\documentclass[twocolumn,floats,showpacs,superscriptaddress,pre]{revtex4}
\usepackage{amsmath,amssymb,bm}
\usepackage{graphics}
\usepackage{epsfig}
\usepackage{graphicx}

\begin{document}

\title{On a New Form of Quantum Mechanics (II)}
\author{Natalia Gorobey, Alexander Lukyanenko}
\email{alex.lukyan@rambler.ru}
\affiliation{Department of Experimental Physics, St. Petersburg State Polytechnical
University, Polytekhnicheskaya 29, 195251, St. Petersburg, Russia}
\author{Inna Lukyanenko}
\email{inna.lukyanen@gmail.com}
\affiliation{Institut f\"{u}r Mathematik, TU Berlin, Strasse des 17 Juni 136, 10623
Berlin, Germany}

\begin{abstract}
The correspondence of a new form of quantum mechanics based on a quantum
version of the action principle, which was proposed earlier, with the
ordinary quantum mechanics is established. New potentialities of the quantum
action principle in the interpretation of quantum mechanics are considered.
\end{abstract}

\maketitle
\date{\today }





\section{\textbf{INTRODUCTION}}

In the work \cite{GL1} a new form of quantum mechanics based on a quantum
version of the action principle was formulated for the first time. The new
formulation becomes more exact in the subsequent work \cite{GL2}. However,
the correspondence of the new framework with the ordinary formulation of
quantum mechanics remained not quite clear. The purpose of the present work
is to fill in this gap. In addition, new potentialities of the quantum action
principle (QAP) in the interpretation of quantum mechanics are considered.

\section{QUANTUM ACTION PRINCIPLE}

Let us begin with the classical action of a non-relativistic particle in a
canonical form:
\begin{equation}
I=\int\limits_{0}^{T}dt\left( p_{k}\overset{\cdot }{x}_{k}-\frac{p^{2}}{2m}%
-U\left( x,t\right) \right)  \label{1}
\end{equation}%
In the new form of canonical quantization procedure proposed in \cite{GL1},%
\cite{GL2}, canonical variables $\left( x_{k},p_{k}\right) $ are represented as
operators in a space of wave functionals $\Psi \left[ x\left( t\right) %
\right] $ as follows:
\begin{eqnarray}
\widehat{x}_{k}\left( t\right) \Psi &\equiv &x_{k}\left( t\right) \Psi ,
\label{2} \\
\widehat{p}_{k}\left( t\right) \Psi &\equiv &\frac{\widetilde{\hbar }}{i}%
\frac{\delta \Psi }{\delta x_{k}\left( t\right) }.  \notag
\end{eqnarray}%
The constant $\widetilde{\hbar }$ is not equal to the ordinary Plank
constant $\hbar $. Its physical dimensionality is $\left[
\widetilde{\hbar }\right] =Joule\cdot s^{2}$. A relationship between two
constants will be introduced here as a central point of the correspondence
between two forms of quantum mecanics. The operators (\ref{2}) are formally
Hermitian with respect to a scalar product in a space of wave functionals:
\begin{equation}
\left( \Psi _{1},\Psi _{2}\right) \equiv \int \prod\limits_{t}d^{3}x\left(
t\right) \overline{\Psi }_{1}\left[ x\left( t\right) \right] \Psi _{2}\left[
x\left( t\right) \right] .  \label{3}
\end{equation}%
The operator representation (\ref{2}) permits us to define an action operator
as follows:
\begin{equation}
\widehat{I}\equiv \int\limits_{0}^{T}dt\left[ \frac{\widetilde{\hbar }}{i}%
\overset{\cdot }{x}_{k}\left( t\right) \frac{\delta }{\delta x_{k}\left(
t\right) }+\frac{\widetilde{\hbar }^{2}}{2m}\frac{\delta ^{2}}{\delta
x^{2}\left( t\right) }-U\left( x\left( t\right) ,t\right) \right]  \label{4}
\end{equation}%
The first term of the integrand (\ref{4}) is non-Hermitian, however, we can overcome this problem
by throwing away the corresponding imaginary parts of
eigenvalues of the action operator. Two remaining terms are formally
Hermitian with respect to the scalar product (\ref{3}).

Let us turn to the formulation of QAP. For the action operator (\ref{4}) we
consider the eigenvalue problem:
\begin{equation}
\widehat{I}\Psi =\lambda \Psi .  \label{5}
\end{equation}%
The statement is that the equation (\ref{5}) is an analog of Schr\"{o}dinger
equation. It is usefull to re-formulate the eigenvalue problem, introducing
for any wave functional $\Psi $ a functional:
\begin{equation}
\Lambda \left[ x\left( t\right) \right] \equiv \frac{\widehat{I}\Psi \left[
x\left( t\right) \right] }{\Psi \left[ x\left( t\right) \right] }.  \label{6}
\end{equation}%
For the wave functional the exponential representation
\begin{equation}
\Psi \left[ x\left( t\right) \right] \equiv \exp \left( \frac{i}{\widetilde{%
\hbar }}S\left[ x\left( t\right) \right] +R\left[ x\left( t\right) \right]
\right)  \label{7}
\end{equation}%
with real functionals $S\left[ x\left( t\right) \right] ,R\left[ x\left(
t\right) \right] $ will be usefull, in particular, for quasi-classical
decomposition of a solution. Substituting (\ref{7}) in (\ref{6}), we obtain:
\begin{eqnarray}
\Lambda \left[ x\right] &=&\Lambda _{Re}\left[ x\right] +i\widetilde{%
\hbar }\Lambda _{Im}\left[ x\right] ,  \label{8} \\
\Lambda _{Re}\left[ x\right] &\equiv &\int\limits_{0}^{T}dt\left\{
\overset{\cdot }{x}_{k}\frac{\delta S}{\delta x_{k}}+\frac{1}{2m}\left[
-\left( \frac{\delta S}{\delta x_{k}}\right) ^{2}\right. \right.  \notag \\
&&\left. \left. +\widetilde{\hbar }^{2}\left( \left( \frac{\delta R}{\delta
x_{k}}\right) ^{2}+\frac{\delta ^{2}R}{\delta x_{k}^{2}}\right) \right]
-U\right\} ,  \label{9} \\
\Lambda _{Im}\left[ x\right] &\equiv &\int\limits_{0}^{T}dt\left[ -%
\overset{\cdot }{x}_{k}\frac{\delta R}{\delta x_{k}}\right.  \notag \\
&&\left. +\frac{1}{2m}\left( 2\frac{\delta S}{\delta x_{k}}\frac{\delta R}{%
\delta x_{k}}+\frac{\delta ^{2}S}{\delta x_{k}^{2}}\right) \right] .
\label{10}
\end{eqnarray}

The necessary condition of equality of the functional (\ref{6}) to an
eigenvalue of the action operator is its independence on internal points $%
x\left( t\right) ,t\in \left( 0,T\right) $ of a trajectory. Only dependence
on the end points $x_{0k},x_{Tk}$, which are fixed at this stage, is admited.
Now analytical properties of the potential $U\left( x,t\right) $ will be
important. We shall assume that this potential is a real-analytical function
of $x$, i.e., it can be represented by a series:
\begin{equation}
U\left( x,t\right) =U_{0}\left( t\right) +U_{1k}\left( t\right) x_{k}+\frac{1%
}{2}U_{2kl}\left( t\right) x_{k}x_{l}+....  \label{11}
\end{equation}%
Then the functionals $S\left[ x\left( t\right) \right] ,R\left[ x\left(
t\right) \right] $ can be represented as functional series:
\begin{eqnarray}
S\left[ x\left( t\right) \right]  &=&\int\limits_{0}^{T}dts\left( x\left(
t\right) ,t\right) ,  \label{12} \\
s\left( x\left( t\right) ,t\right)  &\equiv &s_{1k}\left( t\right)
x_{k}\left( t\right)   \notag \\
&&+\frac{1}{2}s_{2kl}\left( t\right) x_{k}\left( t\right) x_{l}\left(
t\right) +...,  \notag \\
R\left[ x\left( t\right) \right]  &=&\int\limits_{0}^{T}dtr\left( x\left(
t\right) ,t\right) ,  \label{13} \\
r\left( x\left( t\right) ,t\right)  &\equiv &r_{1k}\left( t\right)
x_{k}\left( t\right)   \notag \\
&&+\frac{1}{2}r_{2kl}\left( t\right) x_{k}\left( t\right) x_{l}\left(
t\right) +....  \notag
\end{eqnarray}%
Let us outline that the correspondence with ordinary quantum mechanics will
take place on a narrow class of wave functionals with local in time
functionals (\ref{12}), (\ref{13}). Substituting decompositions (\ref{12}), (%
\ref{13}) in (\ref{9}), (\ref{10}), and integrating by parts of the first terms
under integrals, we obtain:
\begin{eqnarray}
\Lambda _{Re}\left[ x\right] &=&\left. s\left( x\left( t\right)
,t\right) \right\vert _{0}^{T}-\int\limits_{0}^{T}dt\left\{ \overset{\cdot }%
{s}+\frac{1}{2m}\left( \frac{\partial s}{\partial x_{k}}\right) ^{2}+U\right.
\notag \\
&&\left. -\frac{\widetilde{\hbar }^{2}}{2m}\left[ \left( \frac{\partial r}{%
\partial x_{k}}\right) ^{2}+\frac{\partial ^{2}r}{\partial x_{k}^{2}}\right]
\right\} ,  \label{14} \\
\Lambda _{Im}\left[ x\right] &=&-\left. r\left( x\left( t\right)
,t\right) \right\vert _{0}^{T}+\int\limits_{0}^{T}dt\left[ \overset{\cdot }{%
r}\right.  \notag \\
&&+\left. \frac{1}{2m}\left( 2\frac{\partial s}{\partial x_{k}}\frac{%
\partial r}{\partial x_{k}}+\frac{\partial ^{2}s}{\partial x_{k}^{2}}\right) %
\right] .  \label{15}
\end{eqnarray}%
We must put equal to zero all coefficients in front of non-zero degrees of $%
x_{k}\left( t\right) $ in the integrands (\ref{14}), (\ref{15}).
This necessary condition leads to an infinite system of differential
equations for coefficients of the series (\ref{12}) and (\ref{13}):
\begin{eqnarray*}
&&\overset{\cdot }{s}_{1k}+\frac{1}{m}s_{1l}s_{2lk}-\frac{\widetilde{\hbar }%
^{2}}{2m}\left( 2r_{1l}r_{2lk}+r_{3llk}\right) +U_{1k} \\
&=&0,
\end{eqnarray*}
\begin{eqnarray}
&&\overset{\cdot }{s}_{2kl}+\frac{1}{m}\left(
s_{2mk}s_{2ml}+2s_{1m}s_{3mkl}\right)  \notag \\
&&-\frac{\widetilde{\hbar }^{2}}{m}\left(
r_{2mk}r_{2ml}+2r_{1m}r_{3mkl}\right) +U_{2kl}  \label{16} \\
&=&0,  \notag \\
&&...,  \notag
\end{eqnarray}
\begin{eqnarray*}
&&\overset{\cdot }{r}_{1k}+\frac{1}{m}\left(
s_{1l}r_{2lk}+r_{1l}s_{2lk}\right) +\frac{1}{2m}s_{3llk} \\
&=&0,
\end{eqnarray*}
\begin{eqnarray*}
&&\overset{\cdot }{r}_{2kl}+\frac{2}{m}\left(
s_{1m}r_{3mkl}+s_{2mk}r_{2ml}+s_{3mkl}r_{1m}\right) \\
&&+\frac{1}{m}s_{4mmkl} \\
&=&0, \\
&&...,
\end{eqnarray*}%
It is the system that, with two additions, is equivalent to the Schr\"{o}dinger
equation. These additions arise from the coefficients in front of zero
degrees of $x_{k}$ in the integrands (\ref{14}) and (\ref{15}). The first one
appears in the imaginary part (\ref{15}) of the functional (\ref{6}), and it
has to be equal zero:
\begin{equation}
\int\limits_{0}^{T}dt\left( 2s_{1k}r_{1k}+s_{2kk}\right) =0,  \label{17}
\end{equation}%
if the action operator is Hermitian. The remaining part of (\ref{15}) we omit
in accordance with the notion made after the (\ref{4}). The second addition
appears from the real part (\ref{14}) of the functional (\ref{6}):
\begin{equation}
f\left( t\right) \equiv \frac{1}{2m}s_{1k}^{2}-\frac{\widetilde{\hbar }^{2}}{%
2m}\left( r_{1k}^{2}+r_{2kk}\right) +U_{0}.  \label{18}
\end{equation}%
This function is not equal zero. On the one hand, it will be added to the
Hamiltonian operator of the Schr\"{o}dinger theory. This addition does not
change the physical content of the theory. On the other hand, it will be a
part of an action eigenvalue:
\begin{equation}
\lambda =\Lambda _{Re}=\left. s\right\vert
_{0}^{T}+\int\limits_{0}^{T}dtf\left( t\right) ,  \label{19}
\end{equation}

\section{CORRESPONDENCE OF QAP WITH ORDINARY QUANTUM MECHANICS}

To establish this correspondence, let us consider a multiplicative
representation of a wave functional \cite{GL1}. For this purpose, we divide
the interval of time $\left[ 0,T\right] $ into $N$ small parts of an equal length
$\varepsilon =T/N$, and approximate a trajectory $x_{k}\left( t\right) $ by
a piecewise linear function
with vertices $x_{k}\left( t_{n}\right) ,t_{n}=n\varepsilon $;
$x_{k}\left( 0\right) =x_{0k}$, $x_{k}\left( T\right) =x_{Tk}$. Then the
functionals (\ref{12}) and (\ref{13}) can be approximated by corresponding
integral sums:
\begin{equation}
S\left[ x\right] =\sum\limits_{n=1}^{N}\varepsilon s\left(
x_{n},t_{n}\right) ,R\left[ x\right] =\sum\limits_{n=1}^{N}\varepsilon
r\left( x_{n},t_{n}\right) .  \label{20}
\end{equation}%
The central point of the correspondence between two forms of quantum
mechanics is the equality \cite{GL1}:
\begin{equation}
\widetilde{\hbar }=\varepsilon \hbar .  \label{21}
\end{equation}%
Taking into account Eqs. (\ref{20}) and (\ref{21}), the exponential representation of a
wave functional (\ref{7}) can be transformed to the product of wave
functions taken at discrete moments of time:
\begin{eqnarray}
\Psi \left[ x\right] &=&\prod\limits_{n=1}^{N}\psi \left(
x_{n},t_{n}\right) ,\psi \left( x_{n},t_{n}\right) \equiv \exp \chi \left(
x_{n},t_{n}\right) .  \notag \\
\chi \left( x_{n}t_{n}\right) &\equiv &\frac{i}{\hbar }s\left(
x_{n},t_{n}\right) +\varepsilon r\left( x_{n},t_{n}\right)  \label{23}
\end{eqnarray}%
Here and further the product $\varepsilon r$ will be considered as a single
symbol. In this approximation a wave functional $\Psi $\ is a function of
many variables - coordinates of vertices $x_{k}\left( n\right) $ of a broken
line, and its variational derivative has to be replaced by the partial
derivative as follows \cite{GL1}:
\begin{equation}
\frac{\delta \Psi }{\delta x_{k}\left( t_{n}\right) }\equiv \frac{1}{%
\varepsilon }\frac{\partial \Psi }{\partial x_{k}\left( t_{n}\right) }=\frac{%
1}{\varepsilon }\frac{\partial \psi \left( x_{n},t_{n}\right) }{\partial
x_{k}\left( t_{n}\right) }.  \label{24}
\end{equation}%
Then the action operator may be approximated by the differential operator:
\begin{eqnarray}
\widehat{I}\Psi &=&\sum\limits_{n=1}^{N}\varepsilon \left[ \frac{\hbar }{i}%
\frac{x_{k}\left( t_{n}\right) -x_{k}\left( t_{n-1}\right) }{\varepsilon }%
\frac{\partial \Psi }{\partial x_{k}\left( t_{n}\right) }\right.  \notag \\
&&\left. +\frac{\hbar ^{2}}{2m}\frac{\partial ^{2}\Psi }{\partial
x_{k}^{2}\left( t_{n}\right) }-U\left( x_{n},t_{n}\right) \Psi \right] .
\label{25}
\end{eqnarray}%
The first of three parts in the right hand side of Eq. (\ref{25}) can be
transformed as follows:
\begin{eqnarray}
&&\sum\limits_{n=1}^{N}\varepsilon \frac{\hbar }{i}\frac{x_{k}\left(
t_{n}\right) -x_{k}\left( t_{n-1}\right) }{\varepsilon }\frac{\partial \Psi
}{\partial x_{k}\left( t_{n}\right) }  \notag \\
&=&\sum\limits_{n=1}^{N}\varepsilon \frac{\hbar }{i}\frac{x_{k}\left(
t_{n}\right) -x_{k}\left( t_{n-1}\right) }{\varepsilon }\frac{\partial \chi
\left( x_{n},t_{n}\right) }{\partial x_{k}\left( t_{n}\right) }\Psi  \notag
\\
&\cong &\sum\limits_{n=1}^{N}\varepsilon \frac{\hbar }{i}\left[ \frac{\chi
\left( x_{n},t_{n}\right) -\chi \left( x_{n-1},t_{n-1}\right) }{\varepsilon }%
-\frac{\partial \chi \left( x_{n},t_{n}\right) }{\partial t_{n}}\right] \Psi
\notag \\
&=&\left[ \left. \frac{\hbar }{i}\chi \left( x_{n},t_{n}\right) \right\vert
_{n=0}^{n=N}-\sum\limits_{n=1}^{N}\varepsilon \frac{\hbar }{i}\frac{%
\partial \chi \left( x_{n},t_{n}\right) }{\partial t_{n}}\right] \Psi
\label{26}
\end{eqnarray}%
The approximate equality becomes exact in the limit $N\rightarrow \infty $%
, which is supposed. The functional (\ref{7}) in this approximation takes a
form:
\begin{equation}
\Lambda \left[ x\right] =\lambda +\sum\limits_{n=1}^{N}\varepsilon \frac{%
\check{S}ch\psi \left( x_{n},t_{n}\right) }{\psi \left( x_{n},t_{n}\right) },
\label{27}
\end{equation}%
where $\lambda $ is an eigenvalue of the action operator (\ref{25})
approximated as follows:
\begin{equation}
\lambda \equiv s\left( x_{T},T\right) -s\left( x_{0},0\right)
+\sum\limits_{n=1}^{N}\varepsilon f\left( t_{n}\right) ,  \label{28}
\end{equation}%
$f$ is defined by (\ref{18}), and
\begin{equation}
\check{S}ch\psi \equiv i\hbar \overset{\cdot }{\psi }+\frac{\hbar ^{2}}{2m}%
\frac{\partial ^{2}\psi }{\partial x^{2}}-U\psi -f\psi =0  \label{29}
\end{equation}%
is a part of the functional, which depends on non-zero degrees of $%
x_{k}\left( t\right) $. Here we have omitted all imaginary contributions in
accordance with notions made before. Therefore, QAP reduces in this
approximation to the Schr\"{o}dinger equation for a wave function with addition
to the Hamiltonian a function of time defined by (\ref{18}). Let us remember
that this approximation is exact in the limit $N\rightarrow \infty $.

The multiplicative representation of a wave functional (\ref{23}) and its
connection with a solution of Schr\"{o}dinger equation (\ref{29}) gives us a
simple instruction for the probabilistic interpretation of our approach:
a wave functional $\Psi \left[ x\right] $ in the discrete
approximation (\ref{23}) is a complex amplitude of probability of a particle
movement along a broken line between the end points $x_{0k},x_{Tk}$. More
precise, if $\left[ x_{n}-\delta _{n},x_{n}+\delta _{n}\right] $ is an
interval in a neighborhood of the vertice $x_{n}$, then the probability of
particle movement along a broken line from that neighborhood is given by
the expression:
\begin{equation}
P_{\delta }\left[ x\right] =\int\limits_{x_{0}-\delta _{0}}^{x_{0}+\delta
_{0}}d^{3}x_{0}\int\limits_{x_{1}-\delta _{1}}^{x_{1}+\delta
_{1}}d^{3}x_{1}...\int\limits_{x_{T}-\delta _{T}}^{x_{T}+\delta
_{T}}d^{3}x_{T}\left\vert \Psi \left[ x_{n}\right] \right\vert ^{2}.
\label{30}
\end{equation}%
However, this possibility of the probabilistic interpretation is not quite
correct. Any attempt to localize the electron in intermediate points perturbs its posterior movement. Physically correct is the probability of
the electron localization in a neighborhood of the end points:
\begin{eqnarray}
&&P\left[ \left\vert x-x_{0}\right\vert \leq \delta _{0},\left\vert
x-x_{T}\right\vert \leq \delta _{T}\right]  \notag \\
&=&\int\limits_{x_{0}-\delta _{0}}^{x_{0}+\delta
_{0}}d^{3}x_{0}\int\limits_{R^{3}}d^{3}x_{1}...\int%
\limits_{R^{3}}d^{3}x_{N-1}\int\limits_{x_{T}-\delta _{T}}^{x_{T}+\delta
_{T}}d^{3}x_{T}\left\vert \Psi \left[ x_{n}\right] \right\vert ^{2}  \notag
\\
&=&\int\limits_{x_{0}-\delta _{0}}^{x_{0}+\delta _{0}}d^{3}x_{0}\left\vert
\psi \left( x_{0},0\right) \right\vert ^{2}\int\limits_{x_{T}-\delta
_{T}}^{x_{T}+\delta _{T}}d^{3}x_{T}\left\vert \psi \left( x_{T},T\right)
\right\vert ^{2}  \label{31}
\end{eqnarray}%
In the limit $N\rightarrow \infty $ we arrive at a functional integral over a
space of trajectories with partly fixed end points. In the next section
other possibilities of interpretation of the new framework are considered.

\section{QAP IN \textbf{A STRONG FORM}}

The Schr\"{o}dinger equation (\ref{29}), in fact, is a nonlinear equation, in so
far as the function $f\left( t\right) $ , according to (\ref{18}), depends
on $\psi $. Though the physical content of the Schr\"{o}dinger theory does not
changes with addition to the Hamiltonian an arbitrary function of time, this
term is important for the correspondence of QAP with ordinary quantum
mechanics. At the same time, this function enters in the eigenvalue $\lambda
$ of the action operator. Let us focus our attention on the physical meaning
of the eigenvalue $\lambda $. Being a function of a solution of Schr\"{o}%
dinger equation, $\lambda $ may be considered as a generation function for
"observables". Let us introduce for this purpose a probe field $A_{\mu
}\left( x,t\right) $. For instance, it may be an electromagnetic field. Then
$\lambda $ becomes a functional of $A_{\mu }\left( x,t\right) $. It is this
functional that generates currents and their correlators in electrodynamics.
Precisely,
\begin{equation}
\left\langle j^{\mu }\left( x,t\right) \right\rangle \equiv \frac{\delta
\lambda }{\delta A_{\mu }\left( x,t\right) }  \label{32}
\end{equation}%
is a mean value of the electron current density for a given solution of the Schr%
\"{o}dinger equation. Correlators of currents are equal to higher order
variational derivatives. These "observables" give us information about the
electron movement in a "soft" form. However, this form of observation also
works up to the moment when the probe field will perturb the electron
movement.

Finally, we propose more questionable interpretation of eigenvalues of the
action operator, based on a strong form of QAP, which was formulated in \cite%
{GL1},\cite{GL2}. In our framework $\lambda $ is open dependent on phases $%
s\left( x_{0},0\right)$ and $s\left( x_{T},T\right) $ of initial and final wave functions taken at the end points $x_{0k},x_{Tk}$. These phases
are represented by series:
\begin{eqnarray}
s\left( x_{0},0\right) &=&s_{1k}\left( 0\right) x_{0k}+\frac{1}{2}%
s_{2kl}\left( 0\right) x_{0k}x_{0l}+...,  \notag \\
s\left( x_{T},T\right) &=&s_{1k}\left( T\right) x_{Tk}+\frac{1}{2}%
s_{2kl}\left( T\right) x_{Tk}x_{Tl}+....  \label{33}
\end{eqnarray}%
In the strong form of QAP an additional condition of stationarity of an
eigenvalue $\lambda $ with respect to small variations of the coefficients $%
s_{1k}\left( 0\right) ,s_{2kl}\left( 0\right) ,...,r_{1k}\left( 0\right)
,r_{2kl}\left( 0\right) ,...$ was accepted. These coefficients play a role
of initial data for the system of differential equations (\ref{16}), or,
equivalently, for the Schr\"{o}dinger equation (\ref{29}). The stationary
eigenvalue $\lambda _{0}$ becomes a function only of time and coordinates of
the end points: $\lambda _{0}=\lambda _{0}\left( x_{0},x_{T},T\right) $.
What is the physical meaning of this function? In the work \cite{GL2} the
classical limit $\hbar \rightarrow 0$ of this function in the case of a
harmonic oscillator was obtained: it coincides with the classical action
calculated along a stationary trajectory of a particle between the end
points $x_{0k}$ and $x_{Tk}$. We shall suppose that this limit takes place in
general case. Therefore, the stationary eigenvalue $\lambda _{0}$ may be
considered as a quantum analog of the classical action $I\left(
x_{0},x_{T},T\right) $. This statement has no practical meaning if we don't
make the next step. In classical mechanics the action $I\left(
x_{0},x_{T},T\right) $ is a generating function of the canonical
transformation from the initial canonical variables $\left(
x_{0},p_{0}\right) $ to the final ones $\left( x_{T},p_{T}\right) $, which
is defined by the equations (see, for example, \cite{X}):
\begin{equation}
\frac{\partial I\left( x_{0,}x_{T},T\right) }{\partial x_{0k}}=-p_{0k},\frac{%
\partial I\left( x_{0,}x_{T},T\right) }{\partial x_{Tk}}=p_{Tk}.  \label{34}
\end{equation}%
These equations gives us coordinates of the end point as a function of time
and the initial data: $x_{Tk}=x_{Tk}\left( x_{0k},p_{0k},T\right)$. It is a
solution of classical equations of motion. Our proposal is to consider a
quantum analog of Eq. (\ref{34}):
\begin{equation}
\frac{\partial \lambda _{0}\left( x_{0,}x_{T},T\right) }{\partial x_{0k}}%
=-p_{0k},\frac{\partial \lambda _{0}\left( x_{0,}x_{T},T\right) }{\partial
x_{Tk}}=p_{Tk}.  \label{35}
\end{equation}%
If the equations (\ref{35}) have a unique solution for coordinates of the
end point $x_{Tk}=x_{Tk}\left( x_{0k},p_{0k},T\right) $, one can interpret
it as a causal prediction for an electron movement in the framework of a
theory of hidden parameters analogous to the D.Bohm theory \cite{BH}. Here
the initial data $\left( x_{0k},p_{0k}\right) $ form a set of hidden
parameters with a probabilistic measure defined by an initial quantum state
of the system.

\section{\textbf{CONCLUSIONS }}

In conclusion, we showed that the new formulation of quantum mechanics based on a quantum
version of the action principle, taken on a narrow class of exponential wave
functionals, is equivalent to the ordinary Schr\"{o}dinger formulation of
quantum mechanics. This new approach gives new potentialities for interpretation
of quantum mechanics. In addition to the usual probabilistic interpretation
in terms of a wave function (or a wave functional) one obtains a quantum
analog of the classical action, which may be used as a basis of a theory of
hidden parameters.

We thank V. A. Franke and A. V. Goltsev for useful discussions.




\end{document}